\documentclass[aps,prc,preprint,showpacs,superscriptaddress]{revtex4-1}
\usepackage[dvips]{graphicx}
\usepackage{CJK}
\usepackage{xspace}
\usepackage[T1]{fontenc}
\usepackage{mathptmx}
\usepackage{multirow}
\usepackage{pstricks}
\usepackage{psfrag}

\newcommand{\kn}[1]{{\footnotesize (#1)}}   % Kanji-name (smaller font)
\newcommand{\ts}{\textsuperscript}
\newcommand{\etal}{\textit{et al.}\xspace}
\newcommand{\gammaray}{\ensuremath{\gamma}-ray\xspace}

\newcommand{\ioi}{``Island of Inversion''\xspace}
\newcommand{\etwop}{\ensuremath{E(2^+_1)}\xspace}
\newcommand{\beupl}{\ensuremath{B(E2;~0^+_{\mathrm{g.s.}}\rightarrow2^+_1)}\xspace} % long B(E2)
\newcommand{\mevu}{MeV/nucleon\xspace}

\newcommand{\tnaa} {\ensuremath{(5/2^+_1)\rightarrow3/2^{(+)}_{\mathrm{g.s.}}}\xspace} %the first transition in 31na
\newcommand{\tnab} {\ensuremath{(7/2^+_1)\rightarrow(5/2^+_1)}\xspace}                %the second transition in 31na in coincidence
\newcommand{\tenaa}{376(4) keV\xspace} % the first excited state in 31na
\newcommand{\tenab}{787(8) keV\xspace} % the second excited stat in 31na
\newcommand{\tenac}{569(12) keV\xspace} % the first excited state in 32na
\newcommand{\tenada}{447(13) keV\xspace} % the first excited state in 33na from knockout
\newcommand{\tenadb}{476(12) keV\xspace} % the first excited state in 33 na from CC'
\newcommand{\tenad}{467(13) keV\xspace} % average of the above two 

\newcommand{\tenaanokev}{376(4)\xspace} % the first excited state in 31na
\newcommand{\tenabnokev}{787(8)\xspace} % the second excited stat in 31na
\newcommand{\tenacnokev}{569(12)\xspace} % the first excited state in 32na
\newcommand{\tenadnokev}{467(13)\xspace} % average of the above two 

\begin{document}
\begin{CJK}{UTF8}{min}
 
  % 
  % frontmatter
  % 
  \title{Exploring the \ioi by
    in-beam $\gamma$-ray spectroscopy of the neutron-rich sodium isotopes \ts{31,32,33}Na}

  % commands for affiliation
  \newcommand{\ariken}{  \affiliation{RIKEN Nishina Center, Wako, Saitama 351-0198, Japan}}
  \newcommand{\apeking}{ \affiliation{Peking University, Beijing 100871, P.R. China}}
  \newcommand{\atum}{    \affiliation{Physik Department E12, Technische Universit\"at M\"unchen, 85748 Garching, Germany}}
  \newcommand{\alpc}{    \affiliation{LPC-Caen, ENSICAEN, Universit\'e de Caen, CNRS/IN2P3, 14050 Caen cedex, France}}
  \newcommand{\arikkyo}{ \affiliation{Department of Physics, Rikkyo University, Toshima, Tokyo 172-8501, Japan}}
  \newcommand{\auot}{    \affiliation{Department of Physics, University of Tokyo, Bunkyo, Tokyo 113-0033, Japan}}
  \newcommand{\atit}{    \affiliation{Department of Physics, Tokyo Institute of Technology, Meguro, Tokyo 152-8551, Japan}}
  \newcommand{\acns}{    \affiliation{Center for Nuclear Study, The University of Tokyo, RIKEN Campus, Wako, Saitama 351-0198, Japan}}
  \newcommand{\atus}{    \affiliation{Department of Physics, Tokyo University of Science, Noda, Chiba 278-8510, Japan}}
  \newcommand{\asaitama}{\affiliation{Department of Physics, Saitama University, Saitama 338-8570, Japan}}
  \newcommand{\agsi}{    \affiliation{GSI Helmholtzzentrum f\"ur Schwerionenforschung GmbH, 64291 Darmstadt, Germany}}

  \newcommand{\aem}{\email{pieter@ribf.riken.jp}}  % author email

  \author{P.~Doornenbal}         \aem  \ariken             % EMAIL: pieter@ribf.riken.jp
  \author{H.~Scheit}                   \ariken  \apeking   % EMAIL: scheit@ribf.riken.jp
  \author{N.~Kobayashi}                 \atit              % EMAIL: kobayashi.n.aa@m.titech.ac.jp
  % - gamma team first
  \author{N.~Aoi \kn{青井考}}            \ariken             % EMAIL: aoi@riken.jp
  \author{S.~Takeuchi \kn{武内聡}}       \ariken             % EMAIL: takesato@ribf.riken.jp
  \author{K.~Li \kn{李闊昂}}             \ariken  \apeking   % EMAIL: lika@ribf.riken.jp
  \author{E.~Takeshita \kn{竹下英里}}     \ariken             % EMAIL: eri@riken.jp
  \author{Y.~Togano \kn{栂野泰宏}}        \ariken            % EMAIL: togano@ribf.riken.jp
  \author{H.~Wang}                      \ariken \apeking   % EMAIL: wanghe@ribf.riken.jp
  % - from here purely alphabetical
  %\author{H.~Baba \kn{馬場秀忠}}          \ariken            % EMAIL: baba@ribf.riken.jp
  \author{S.~Deguchi}                   \atit              % EMAIL: deguchi.s.aa@m.titech.ac.jp
  %\author{N.~Fukuda \kn{福田直樹}}        \ariken            % EMAIL: nfukuda@riken.jp
  %\author{H.~Geissel}                   \agsi              % EMAIL: H.Geissel@gsi.de
  %\author{R.~Gernh\"auser}              \atum              % EMAIL: roman.gernhaeuser@ph.tum.de
  %\author{J.~Gibelin}                   \alpc              % EMAIL: gibelin@lpccaen.in2p3.fr
  %\author{I.~Hachiuma}                  \asaitama          % EMAIL: hachiuma@ne.phy.saitama-u.ac.jp
  %\author{Y.~Hara}                      \arikkyo           % EMAIL: 08la015t@rikkyo.ac.jp
  %\author{C.~Hinke}                     \atum              % EMAIL: christoph.hinke@ph.tum.de
  %\author{N.~Inabe \kn{稲辺尚人}}         \ariken            % EMAIL: inabe@ribf.riken.jp
  %\author{K.~Itahashi \kn{板橋健太}}      \ariken            % EMAIL: itahashi@riken.jp
  %\author{S.~Itoh}                      \auot              % EMAIL: itoh@nucl.phys.s.u-tokyo.ac.jp
  %\author{D.~Kameda \kn{亀田大輔}}        \ariken            % EMAIL: kameda@ribf.riken.jp
  %\author{S.~Kanno \kn{菅野祥子}}         \ariken            % EMAIL: shoko@riken.jp
  \author{Y.~Kawada}                    \atit              % EMAIL: kawada.y.aa@m.titech.ac.jp
  \author{Y.~Kondo \kn{近藤洋介}}         \ariken            % EMAIL: kondo@ribf.riken.jp
  %\author{R.~Kr\"ucken}                 \atum              % EMAIL: reiner.kruecken@ph.tum.de
  %\author{T.~Kubo \kn{久保敏幸}}          \ariken            % EMAIL: kubo@ribf.riken.jp
  %\author{T.~Kuboki}                    \asaitama          % EMAIL: kuboki@ne.phy.saitama-u.ac.jp
  %\author{K.~Kusaka \kn{日下健祐}}        \ariken            % EMAIL: kkusaka@riken.jp
  %\author{M.~Lantz}                     \ariken            % EMAIL: lantz@ribf.riken.jp
  %\author{S.~Michimasa}                 \acns              % EMAIL: mitimasa@cns.s.u-tokyo.ac.jp
  \author{T.~Motobayashi \kn{本林透}}     \ariken            % EMAIL: motobaya@riken.jp
  \author{T.~Nakamura \kn{中村隆司}}      \atit              % EMAIL: nakamura@ap.titech.ac.jp
  %\author{T.~Nakao}                     \auot              % EMAIL: nakao@nucl.phys.s.u-tokyo.ac.jp
  %\author{K.~Namihira}                  \asaitama          % EMAIL: namihira@ne.phy.saitama-u.ac.jp
  %\author{S.~Nishimura \kn{西村俊二}}     \ariken            % EMAIL: nishimu@riken.jp
  %\author{T.~Ohnishi \kn{大西哲哉}}       \ariken            % EMAIL: oonishi@ribf.riken.jp
  %\author{M.~Ohtake \kn{大竹政雄}}        \ariken            % EMAIL: mohtake@riken.jp
  %\author{N.A.~Orr}                     \alpc              % EMAIL: orr@lpccaen.in2p3.fr
  %\author{H.~Otsu \kn{大津秀暁}}          \ariken            % EMAIL: otsu@ribf.riken.jp
  %\author{K.~Ozeki \kn{大関和貴}}         \ariken            % EMAIL: k_ozeki@riken.jp
  \author{Y.~Satou}                     \atit              % EMAIL: satou@phys.titech.ac.jp
  %\author{S.~Shimoura}                  \acns              % EMAIL: shimoura@cns.s.u-tokyo.ac.jp
  %\author{T.~Sumikama \kn{炭竃聡之}}      \atus              % EMAIL: sumikama@ph.noda.tus.ac.jp
  %\author{M.~Takechi \kn{武智麻耶}}       \ariken            % EMAIL: takechi@ribf.riken.jp
  %\author{H.~Takeda \kn{竹田浩之}}        \ariken            % EMAIL: takeda@ribf.riken.jp
  \author{K.~N.~Tanaka}                 \atit              % EMAIL: tanaka@mail.nucl.ap.titech.ac.jp
  %\author{K.~Tanaka \kn{田中鐘信}}        \ariken            % EMAIL: ktanaka@riken.jp
  %\author{M.~Winkler}                   \agsi              % EMAIL: M.Winkler@gsi.de
  %\author{Y.~Yanagisawa \kn{柳澤善行}}    \ariken            % EMAIL: yanagisa@riken.jp
  %\author{K.~Yoneda \kn{米田健一郎}}      \ariken            % EMAIL: kyoneda@riken.jp
  %\author{A.~Yoshida \kn{吉田敦}}        \ariken            % EMAIL: ayoshida@ribf.riken.jp
  %\author{K.~Yoshida \kn{吉田光一}}       \ariken            % EMAIL: yoshida@ribf.riken.jp
  \author{H.~Sakurai \kn{櫻井博儀}}       \ariken            % EMAIL: sakurai@ribf.riken.jp

  \date{\today}

  %_______________________________________________
  % 
  % abstract
  % 
  \begin{abstract}
    The structure of the neutron-rich sodium isotopes \ts{31,32,33}Na was investigated by means of in-beam
    \gammaray spectroscopy following one-neutron knockout and inelastic scattering of radioactive beams 
    provided by the RIKEN Radioactive Ion Beam Factory.  
    The secondary beams were selected and separated by the fragment separator BigRIPS and incident at $\approx$~240~\mevu on
    a natural carbon (secondary) target, which was surrounded by the DALI2 array to detect coincident de-excitation $gamma$ rays.
    Scattered particles were identified by the spectrometer ZeroDegree. 
    In \ts{31}Na, a new decay $gamma$ ray was observed in coincidence with the known \tnaa transition, while for 
    \ts{32,33}Na excited states are reported for the first time. From a comparison to
    state-of-the-art shell-model calculations it is concluded that the newly observed excited state in  
    \ts{31}Na belongs to a rotational band formed by a 2$p$2$h$ intruder configuration within the 
    ``Island of Inversion.''
  \end{abstract}

  \maketitle
\end{CJK}

%_________________________________________________
% 
% Introduction
% 

% general Introduction about the Na isotopes.
%\section{Introduction}

%historical
In the traditional perception of the shell model the magic numbers, which were first reproduced
correctly for stable nuclei by Mayer and Jensen \cite{haxel:1949,mayer:1949} and define the shell closures,
are recognized to be valid globally across the Segr\'e chart. 
Evidence for a sudden, unanticipated change of shell structure in the neutron-rich sodium isotopes
came from anomalously low masses measured for the isotopes \ts{31}Na and \ts{32}Na~\cite{thibault:1975}. 
These increased binding energies were seen as an indication for the quenching of the $N=20$ magic number 
in this region of the Segr\'e chart~\cite{campi:1975}. 

%Theory
% short introduction of the ioi, and reason for the reduced gap.
On the theoretical side, the shell-model study by Warburton \etal~\cite{warburton:1990} identified 
the nine nuclei with $Z=10$--12 and $N=20$--22 as a region in which  
$\nu(sd)^{-2}(fp)^2$ (2$\hbar\omega$) intruder configurations are more tightly bound than 
the normal (0$\hbar\omega$) configurations, making them the ground states (g.s.). This region is since then 
referred to as the \ioi. The inversion of the configurations 
can be attributed to a narrowing of the $N=20$ neutron shell gap between the $d_{3/2}$ and $f_{7/2}$ 
orbitals from a lack of protons in the $d_{5/2}$ shell in neutron-rich nuclei~\cite{otsuka:2001}.

% Experimental situation:
Further experimental evidence for the erosion of the $N=20$ shell closure is given by
the small \etwop energy and large \beupl value  of 
\ts{32}Mg~\cite{detraz:1979,motobayashi:1995} and meanwhile a multitude of experiments have
addressed the evolution of the $N=20$ shell near the 
\ioi. Despite this exceptional attraction, because of experimental restrictions,
only for the even-$Z$ isotopes (Ne, Mg) information 
on excited states has been obtained
from in-beam \gammaray spectroscopy beyond 
$N=20$~\cite{iwasaki:2001,yoneda:2001,pritychenko:2002,church:2005,elekes:2006,gade:2007,doornenbal:2009a}.

Concerning the chain of the odd-$Z$ sodium isotopes, the experimental knowledge is very 
limited. While the ground-state spin is firmly 
established up to $N=20$~\cite{huber:1978,keim:2000}, 
for heavier isotopes, only for certain nuclei the parity and ranges of spin $J$ have been 
experimentally established. For instance for \ts{32}Na, $J\le 4^{-}$ has been deduced from
$\beta$-delayed neutron emission into \ts{31}Mg~\cite{klotz:1993}.
Information on excited states
has been merely extended to the $N=20$ nucleus \ts{31}Na by means of  
intermediate-energy Coulomb excitation~\cite{pritychenko:2001} and inelastic scattering 
on a liquid hydrogen target~\cite{elekes:2006}.
In both cases, the \tnaa transition was observed at energies of 350(20) and 370(12) keV, respectively.

\begin{figure}
  \centering
  \includegraphics[width=8.3cm]{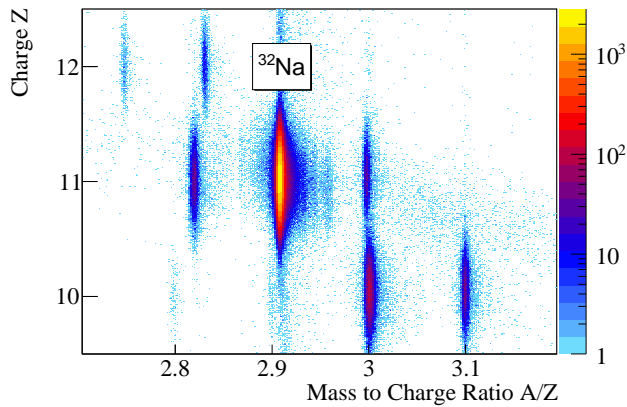}
  \caption {
    (Color online) Particle identification before the secondary 
    target for a BigRIPS setting in which \ts{32}Na
    was the main component of the cocktail beam.
  }
  \label{fig:pid-brips}
\end{figure}

%_________________________________________________
%
% Experimental setup
%
%\section{Experiment}
%The BigRIPS spectrometer:
Here, we report on the observation of excited states in the sodium isotopes with $N$=20--22 (\ts{31,32,33}Na) following
inelastic scattering and one-neutron removal reactions on a natural carbon target at $\approx$~240~\mevu.
The experiment was performed at the Radioactive Ion Beam Factory (RIBF) \cite{yano:2007}, operated by the 
RIKEN Nishina Center and the Center for Nuclear Study, University of Tokyo.
A high-intensity \ts{48}Ca beam at 345~\mevu was incident on a 20-mm thick rotating Be
target \cite{yoshida:2008} at the focus F0 of the fragment separator BigRIPS~\cite{kubo:2003}.
%\footnote{See \cite{sakurai:2008} for the location of the BigRIPS focal points.}. 
From the emerging projectile fragmentation products the neutron-rich
sodium isotopes with mass numbers $A$~=~31--34 were selected and separated using the 
$B\rho$--$\Delta E$--$B\rho$ method in the first stage of BigRIPS 
with the aid of an achromatic aluminum energy degrader of 15-mm thickness 
located at the dispersive focus F1. In the second stage of BigRIPS (from focus F3 to focus F7), the transmitted
fragmentation products were identified event-by-event employing the $\Delta E$--$B\rho$--velocity method.
The energy loss $\Delta E$ was measured by means of an ionization chamber located at
the focus F7 of BigRIPS. The magnetic rigidity $B\rho$ was obtained by position measurements with
parallel plate avalanche counters~\cite{ohnishi:2008} at   
the dispersive focus F5. Two thin plastic scintillators of 1- and 3-mm thickness were placed
at F3 and F7 to measure the time-of-flight (TOF). The path length between both detectors
was 47~m. Figure~\ref{fig:pid-brips} presents an exemplary particle identification plot, showing
clear separation between different nuclides in charge $Z$ and mass to charge ratio $A/Z$. 

%Describe the settings applied to BigRIPS and inform about the resolution:
Three settings with different $B\rho$ values were applied to the BigRIPS fragment separator 
for the study of \ts{31}Na, \ts{32}Na, and \ts{33}Na by one-neutron knockout reactions
and inelastic scattering.
After the selection and identification with BigRIPS, which 
was operated in its full momentum acceptance mode of $\Delta p/p =\pm3$\%,
the secondary beams were incident on a carbon target with 
a 2.54 g/cm\ts{2} thickness and a diameter of 30~mm at the focus F8.
The energy at midtarget varied from $\sim$230~\mevu to $\sim$250~\mevu for the different 
sodium isotopes. The energy loss 
in the secondary target amounted to $\sim$14\% of the incident beam energy.

The DALI2 array \cite{takeuchi:2003}, a NaI(Tl) based \gammaray spectrometer  
consisting of 180 individual crystals, surrounded the secondary target for \gammaray emission angles ranging
from $\vartheta_{\gamma} =11^\circ$ to $\vartheta_{\gamma} = 147^\circ$ in the laboratory system. A full energy peak 
efficiency of 15\% at a \gammaray energy of 1332.5 keV
was measured with a stationary source, in accordance with our GEANT4~\cite{agostinelli:2003} simulations.
No add-back was performed.

% After the secondary target
% ZDS
Reaction products emerging from the secondary target were selected and identified
using the $\Delta E$--$B\rho$--TOF method on an event-by-event basis by the 
spectrometer ZeroDegree~\cite{mizoi:2005,sakurai:2008} from focus F8 to focus F11.
The angular and momentum acceptances were $\sim$80~$\times$~60 mrad\ts{2} and $\pm4$\%, respectively.
The energy loss $\Delta E$ was obtained from an ionization chamber placed at the final focus F11,
the magnetic rigidity $B\rho$ was measured with parallel plate avalanche counters placed at the dispersive foci F9 and F10,
and the TOF was detected between two thin plastic scintillators of 1-mm thickness mounted at F8 and F11 with a 
flight path length of 37~m. The difference in velocity $\Delta\beta$  
before and after the secondary target was used for the selection of particles passing solely through the secondary target.
Particles passing through the target frame or missing the target, in total less than 2 \% of the 
secondary beam intensities, could be clearly identified.

In the present work, excited states in \ts{31-33}Na, populated in one-neutron knockout and inelastic
scattering, are reported. As the results were obtained ``parasitic''
to main experiments~\cite{doornenbal:2009a,nakamura:2009}, 
the momentum distributions of the one-neutron knockout and scattered sodium isotopes were 
not centered in the spectrometer ZeroDegree and
cropped to a large extent; the overall transmissions were much lower than  
$90 \%$.

% 
% Doppler corrected gamma-ray spectra.
% 
% Go by nuclei, first 31Na

\begin{figure}
  \centering
  \includegraphics[width=8.3cm]{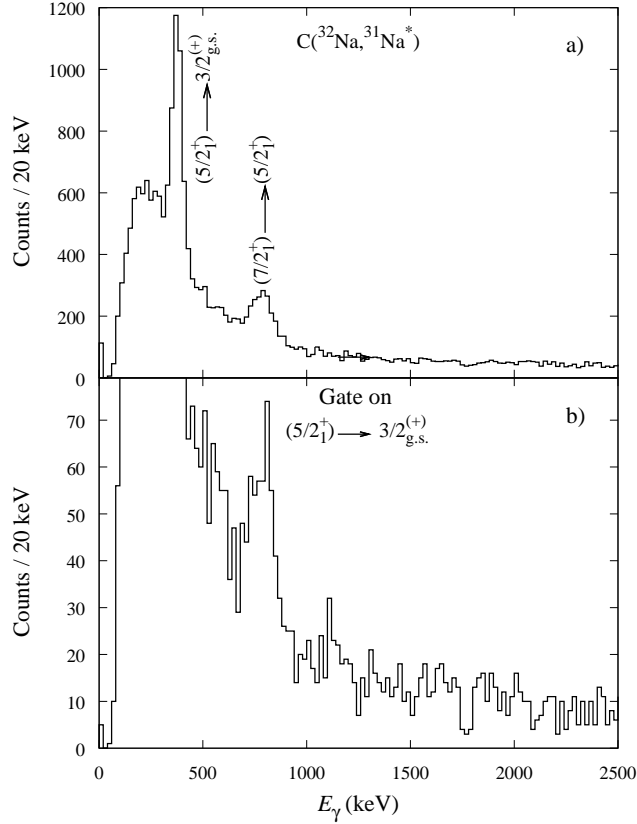}
  \caption  {
    The top panel (a) displays the
    Doppler corrected \gammaray energy spectrum in coincidence ($\pm5$ ns) with 
    the one-neutron removal from \ts{32}Na. In the bottom panel (b), a \gammaray energy
    cut between 325 and 425 keV on the \tnaa transition in \ts{31}Na was applied in addition for events
    of a \gammaray fold greater than one.
  }
  \label{fig:spec-31na}
\end{figure}

Gamma rays emitted from the fast moving reaction products ($\beta \approx 0.6$) 
were Doppler corrected taking into account the lifetimes of the excited states, 
as delineated in Ref.~\cite{doornenbal:2010}. The uncertainty of the lifetime was included
in the uncertainty of the reported \gammaray transition energies.
Figure~\ref{fig:spec-31na} displays the Doppler corrected \gammaray energy spectra 
for the one-neutron removal reactions from \ts{32}Na.
Two distinct transitions are visible at \tenaa and at \tenab. 
The former is in agreement with previous
observations of 350(20)~keV~\cite{pritychenko:2001} and 370(12)~keV~\cite{elekes:2006}
and is generally interpreted to be the \tnaa transition.
Applying a gate ranging from 325 to 425 keV on this transition shows that
both observed decays are in coincidence, as illustrated in the lower panel of Fig.~\ref{fig:spec-31na}.
From the observed coincidence and from the later discussed comparison to shell-model
calculations, we concluded that the observed transition at \tenab is the \tnab decay.
No indication for a direct 
transition from the ($7/2^{+}_1$) state to the ground state was found. This is consistent
with the shell-model prediction of Ref.~\cite{pritychenko:2001} for a strong
$B(M1;7/2^{+}_1\rightarrow 5/2^{+}_1)$ causing the ($7/2^{+}_1$) state to decay with an intensity of 
$95\%$ into the ($5/2^{+}_1$) state. From the intensity ratio of the two transitions it follows
that 61(6)~\% of the \tnaa transition are because of feeding from the ($7/2^{+}_1$) state. 
 
\begin{figure}
  \centering
  \includegraphics[width=8.3cm]{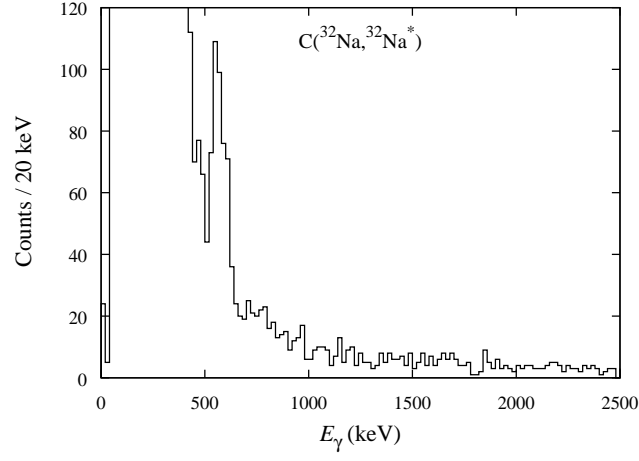}
  \caption  {
    Doppler corrected \gammaray energy spectrum in coincidence ($\pm5$ ns) with 
    inelastic scattering of \ts{32}Na. The \gammaray emission 
    angle was restricted to $\vartheta_\gamma < 90^\circ$.
  }
  \label{fig:spec-32na}
\end{figure}
   
For \ts{32}Na, a \gammaray decay was found at \tenac after inelastic scattering. It is shown in
Fig.~\ref{fig:spec-32na}.

\begin{figure}
  \centering
  \includegraphics[width=8.3cm]{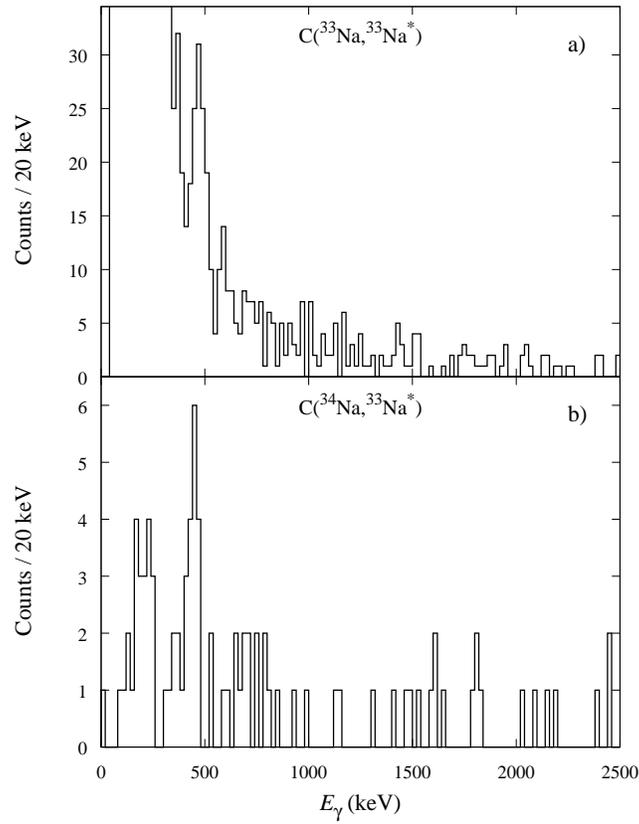}
  \caption  {
    Doppler corrected \gammaray energy spectra in coincidence ($\pm5$ ns) with 
    inelastic scattering of \ts{33}Na (a) and one-neutron removal
    (b). For the former the \gammaray emission 
    angle was restricted to $\vartheta_\gamma < 55^\circ$.
  }
  \label{fig:spec-33na}
\end{figure}

Excited states of \ts{33}Na were detected from inelastic scattering and one-neutron
knockout of \ts{34}Na. Gamma-ray transitions were observed at \tenadb for the former
and at \tenada for the latter case, respectively. As we assume that the two observed peaks 
belong to the same transition, it leads to a combined value of \tenad. 
The Doppler corrected \gammaray spectra are 
shown in Fig.~\ref{fig:spec-33na}. 

% Now Discussing the quality of the spectra
While the one-neutron removal reactions exhibited very low background,
the Doppler corrected \gammaray energy spectra for inelastically scattered \ts{32}Na and \ts{33}Na, 
shown in Fig.~\ref{fig:spec-32na} and Fig.~\ref{fig:spec-33na}(a), respectively, were 
dominated by an exponentially declining distribution that likely originated from atomic processes
during the slowing down of the fragments in the secondary target.  
%The background's energy (angular) distribution was isotopric in the laboratory frame. 
The main component of this background originated from the stationary target
and was thus not Doppler shifted in the laboratory system.
%~\cite{schubert:1994}
Therefore, the observation limit for
low-energy \gammaray transitions shifted as a function of the \gammaray emission angle $\vartheta_\gamma$
with respect to the velocity vector of the emitting nucleus.
%Explain why only low theta angles are usefull for inelastic scattered particles.
Gamma rays emitted toward backward (forward) angles 
are Doppler shifted to lower (higher) energies in the laboratory system.
Thus, low energy \gammaray
transitions are best separated from the background at
forward $\vartheta_\gamma$ angles, when their energy is Doppler shifted to values above
the atomic background. For scattered \ts{32}Na particles an angle cut of $\vartheta_\gamma<90^{\circ}$
was applied; for \ts{33}Na the cut was set accordingly to $\vartheta_\gamma<55^{\circ}$
because of the lower excitation energy at \tenadb.  
Gamma-ray transition energies below $\sim$400~keV could not be measured
in the inelastic channels, even for the 
DALI2 detectors with the lowest $\vartheta_\gamma$ angles. 

\begin{table} 
\caption{\label{tab:results}
 Summary of observed \gammaray transition energies in this work for the isotopes \ts{31-33}Na and proposed spin
 and parity assignments. For \ts{31}Na, the results of Refs.~\cite{elekes:2006,pritychenko:2001}
 are shown for comparison.
}
\begin{ruledtabular}
\begin{tabular}{ccccc}
Isotope & Transition & \multicolumn{3}{c}{Experimental Transition Energies (keV)}\\
        & $J^{\pi}_i \rightarrow J^{\pi}_{f}$ &  This work & Ref.~\cite{elekes:2006} & Ref.~\cite{pritychenko:2001}\\
\hline
\ts{31}Na &\tnaa&\tenaanokev\phantom{0} & 370(12) & 350(20)\\
\ts{31}Na &\tnab&\tenabnokev\phantom{0} & & \\
\ts{32}Na &     &\tenacnokev & & \\
\multirow{2}{*}{\ts{33}Na} & $(5/2_1^+,3/2_1^+)\rightarrow$&\multirow{2}{*}{\tenadnokev} & & \\
  & $(3/2_{\textrm{g.s.}}^+,5/2_{\textrm{g.s.}}^+)$&&
%\hline
\end{tabular}
\end{ruledtabular}
\end{table}

%_________________________________________________ 
% Discussion
%  
%
% Dicsuss only 31Na. What to do with 32,33 Na?
% For 32Na a discussion is difficult, 33Na would require shell model calculations.
% Leave them completely out of the paper?
We will now turn to the discussion part of our experimental results, which are summarized 
in Table~\ref{tab:results}.
% and will emphasize on 
%\ts{31}Na for lack of experimental information on \ts{32,33}Na. 
The low-lying levels of \ts{31}Na have been predicted in previous 
works via shell-model calculations~\cite{pritychenko:2001,caurier:2001,utsuno:2002}.
In the case of Pritychenko \etal, the neutron configuration was restricted to a pure 2$p$--2$h$ (2$\hbar\omega$)
intruder configuration with allowed configurations of $(0d_{5/2})^6$ $(0d_{3/2},s_{1/2})^4$ $(0f_{7/2},1p_{3/2})^2$
for neutrons and of $0d_{3/2}^2$ and $0d_{5/2}^2(1s_{1/2},0d_{3/2})$ for protons, respectively~\cite{pritychenko:2001}.
Utsuno \etal performed Monte Carlo shell-model calculations, allowing for the use of a much
wider model space (the entire $sd$ shell and the $0f_{7/2}$ and the $1p_{3/2}$ orbits) and an unrestricted
mixing of normal and intruder configurations~\cite{utsuno:2002,utsuno:1999}.
The valence space by Caurier \etal included the 
full $sd$ shell for protons in a calculation of the normal configuration and in addition 
the full $pf$ shell for neutrons in a 2$\hbar\omega$ intruder configuration calculation, but
did not allow for any configuration mixing~\cite{caurier:2001,caurier:1998}.

\begin{figure}
  \centering  
  \includegraphics[width=8.3cm]{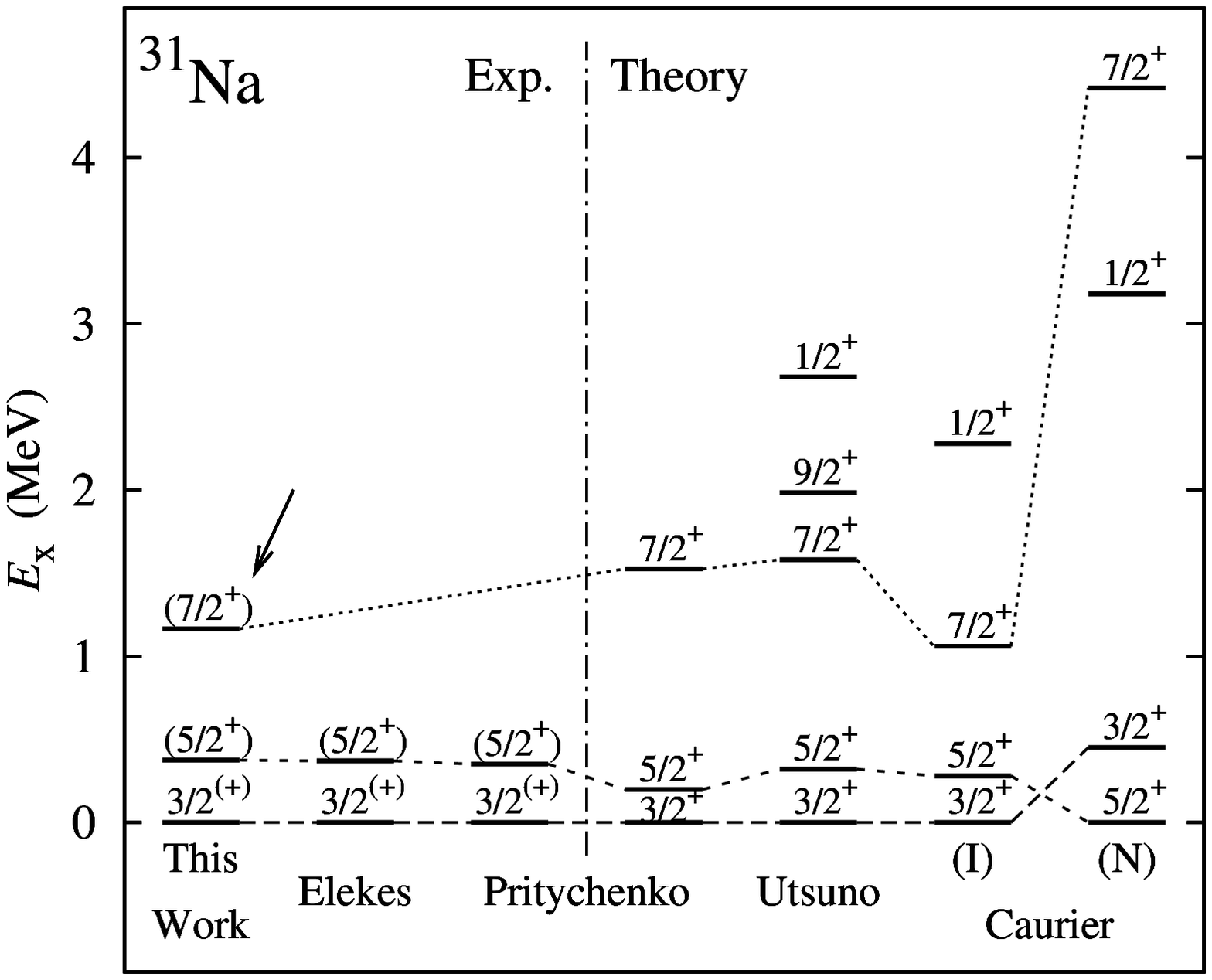}
  \caption {
    Comparison of experimental excitation energies of \ts{31}Na obtained in this
    work, by Elekes~\etal~\cite{elekes:2006}, and by Pritychenko \etal~\cite{pritychenko:2001},  
    with shell-model results by the latter, by
    Utsuno \etal~\cite{utsuno:2002}, and by Caurier \etal~\cite{caurier:2001}. In the latter case,
    normal (N) and intruder (I) configurations are shown. The connecting
    lines between the $3/2^{+}$ (long-dashed line), $5/2^{+}$ (short-dashed line), and $7/2^{+}$ (dotted line)
    levels are drawn to guide the eye and the newly observed state is indicated by the arrow.
    The vertical, dash-dotted line separates the experimental results from the shell-model calculations.
  }
  \label{fig:31nalevel}
\end{figure}

Figure~\ref{fig:31nalevel} compares the above-mentioned shell-model calculations with the experimental data.
Good overall agreement is achieved only for the intruder calculations. On the other hand, the  
normal configuration by Caurier \etal not only predicts the wrong ground-state spin, 
but puts the $7/2^{+}$ level above 4 MeV, at variance with our observations. 

For the odd-odd nucleus \ts{32}Na, several states very close in energy below 200~keV have been
predicted~\cite{poves:1994}. 
As associated low-energy transitions were below our observation limit, no
further conclusions could be drawn on the nature of the transition energy at \tenac.  

\begin{figure} 
  %\centering  
  \includegraphics[width=8.3cm]{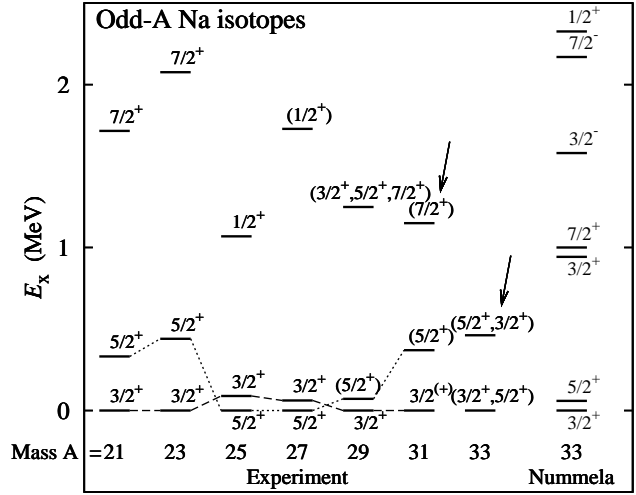}
  \caption {
    Excitation energy of the first and second excited states in particle-bound odd-$A$ sodium isotopes.
    The data are from this work (indicated by the arrows) 
    and~Refs.~\cite{huber:1978,endt:1990,cooper:2002,tripathi:2005,tripathi:2006}.
    The $3/2^{+}$ ($5/2^{+}$) states are connected by dashed (dotted) lines to guide the eye.
    In addition, the shell-model calculations for \ts{33}Na by Nummela \etal~\cite{nummela:2001a} 
    are displayed on the right-hand side. 
  }
  \label{fig:nasystematics}
\end{figure}

To investigate the nature of the \gammaray transition in \ts{33}Na, 
it can be compared to the known low-lying level systematics of particle-bound odd-$A$ sodium
isotopes, displayed in Fig.~\ref{fig:nasystematics}. Shown are all known ground, 
first, and second excited states. 
The two lowest energy states have either spin $3/2^{+}$ or
$5/2^{+}$ and all second excited states lie well above 1 MeV. 
Our experimental setup was insensitive to transition energies below 200 keV for the knockout channels. 
If the observed transition energy at \tenad was a decay from the second to
the first excited state, the level energy would be well below 700 keV, unlike all
other Na isotopes. We therefore propose that the observed transition feeds to the 
ground state. Furthermore, the spin-parity systematics
of the Na isotopes, suggest a $5/2^{+}_1$ level transition to the $3/2^{+}$ ground 
state or vice-versa. 

Shell-model calculations have been performed by Nummela \etal~\cite{nummela:2001a}
for \ts{33}Na that see the ground and first excited state in a $0p0h$ normal configuration.
These calculations, shown on the right-hand side in Fig.~\ref{fig:nasystematics}, agree indeed
with the spin-parity systematics for the ground and first excited state.
However, the levels are almost degenerate in energy, being only 59 keV apart,
which is in contrast to our experimental results of \tenad. Furthermore, transitions
from the $7/2^+_1$ or $3/2^{+}_2$ levels to the  $5/2^+_1$ or  $3/2^{+}_{\mathrm{g.s}}$
levels are predicted at around 1~MeV, which do not agree with our observed transition
energy as well.

%_________________________________________________
% Summary   
In conclusion, we have observed a \gammaray transition
in coincidence to the known \tnaa transition in \ts{31}Na. From a comparison
to state-of-the-art shell-model calculations we concluded that this \tnab transition belongs to an
intruder configuration within the \ioi. In addition, 
for \ts{32}Na and \ts{33}Na \gammaray transitions have been observed for the first time.
With the recent experimental progress owed to the commissioning of the new-generation
facility RIBF, in-beam \gammaray spectroscopy may now be applied to a much wider region of the
\ioi and possibly even beyond its neutron-rich border lines. This in turn, 
necessitates further theoretical fundamental examinations on their exact locations. Finally, we
would like to remark that with the new-generation facility RIBF having gone online,
a new-generation \gammaray spectrometer~\cite{doornenbal:2009}, currently under development, is required
to optimally exploit the vast potential for in-beam \gammaray spectroscopy of fast 
radioactive nuclear beams.   

\begin{acknowledgments}
  We thank the RIKEN Nishina Center staff for 
  providing the high-intensity \ts{48}Ca beam. The measurements were performed in parts parallel
  to other experiments. We would therefore like to thank T.~Oh\-tsubo and his group for
  allowing us to use the BigRIPS spectrometer in the ``Yakitori'' mode.
  P.D. acknowledges the financial support from the Japan
  Society for the Promotion of Science.
  %This work was in part supported by the DFG cluster of excellence {\it Origin and Structure of the Universe}.
\end{acknowledgments}

%  
% references
% 
\bibliography{cb_a1}   % most articles can be referenced as \cite{last-name:year}

%Merlin.mbs v4.21 2009-07-09.
\begin{thebibliography}{10}%
\makeatletter
\providecommand \@ifxundefined [1]{%
 \ifx #1\undefined \expandafter \@firstoftwo
 \else \expandafter \@secondoftwo
\fi
}%
\providecommand \@ifnum [1]{%
 \ifnum #1\expandafter \@firstoftwo
 \else \expandafter \@secondoftwo
\fi
}%
\providecommand \enquote [1]{``#1''}%
\providecommand \bibnamefont  [1]{#1}%
\providecommand \bibfnamefont [1]{#1}%
\providecommand \citenamefont [1]{#1}%
\providecommand\href[0]{\@sanitize\@href}%
\providecommand\@href[1]{\endgroup\@@startlink{#1}\endgroup\@@href}%
\providecommand\@@href[1]{#1\@@endlink}%
\providecommand \@sanitize [0]{\begingroup\catcode`\&12\catcode`\#12\relax}%
\@ifxundefined \pdfoutput {\@firstoftwo}{%
 \@ifnum{\z@=\pdfoutput}{\@firstoftwo}{\@secondoftwo}%
}{%
 \providecommand\@@startlink[1]{\leavevmode\special{html:<a href="#1">}}%
 \providecommand\@@endlink[0]{\special{html:</a>}}%
}{%
 \providecommand\@@startlink[1]{%
  \leavevmode
  \pdfstartlink
   attr{/Border[0 0 1 ]/H/I/C[0 1 1]}%
   user{/Subtype/Link/A<</Type/Action/S/URI/URI(#1)>>}%
  \relax
 }%
 \providecommand\@@endlink[0]{\pdfendlink}%
}%
\providecommand \url  [0]{\begingroup\@sanitize \@url }%
\providecommand \@url [1]{\endgroup\@href {#1}{\urlprefix}}%
\providecommand \urlprefix [0]{URL }%
\providecommand \Eprint[0]{\href }%
\@ifxundefined \urlstyle {%
  \providecommand \doi [1]{doi:\discretionary{}{}{}#1}%
}{%
  \providecommand \doi [0]{doi:\discretionary{}{}{}\begingroup
  \urlstyle{rm}\Url }%
}%
\providecommand \doibase [0]{http://dx.doi.org/}%
\providecommand \Doi[1]{\href{\doibase#1}}%
\providecommand \bibAnnote [3]{%
  \BibitemShut{#1}%
  \begin{quotation}\noindent
    \textsc{Key:}\ #2\\\textsc{Annotation:}\ #3%
  \end{quotation}%
}%
\providecommand \bibAnnoteFile [2]{%
  \IfFileExists{#2}{\bibAnnote {#1} {#2} {\input{#2}}}{}%
}%
\providecommand \typeout [0]{\immediate \write \m@ne }%
\providecommand \selectlanguage [0]{\@gobble}%
\providecommand \bibinfo [0]{\@secondoftwo}%
\providecommand \bibfield [0]{\@secondoftwo}%
\providecommand \translation [1]{[#1]}%
\providecommand \BibitemOpen[0]{}%
\providecommand \bibitemStop [0]{}%
\providecommand \bibitemNoStop [0]{.\EOS\space}%
\providecommand \EOS [0]{\spacefactor3000\relax}%
\providecommand \BibitemShut [1]{\csname bibitem#1\endcsname}%
%</preamble>
\bibitem{haxel:1949}%
  \BibitemOpen
  \bibfield{author}{%
  \bibinfo {author} {\bibfnamefont{O.}~\bibnamefont{Haxel}} \emph{et~al.},\ }%
  \bibfield{journal}{%
  \Doi{10.1103/PhysRev.75.1766.2}{\bibinfo {journal} {Phys. Rev.}}\ }%
  \textbf{\bibinfo {volume} {75}},\ \bibinfo {pages} {1766} (\bibinfo {year}
  {1949})%
  \bibAnnoteFile{NoStop}{haxel:1949}%
\bibitem{mayer:1949}%
  \BibitemOpen
  \bibfield{author}{%
  \bibinfo {author} {\bibfnamefont{M.}~\bibnamefont{{Goeppert Mayer}}},\ }%
  \bibfield{journal}{%
  \Doi{10.1103/PhysRev.75.1969}{\bibinfo {journal} {Phys. Rev.}}\ }%
  \textbf{\bibinfo {volume} {75}},\ \bibinfo {pages} {1969} (\bibinfo {year}
  {1949})%
  \bibAnnoteFile{NoStop}{mayer:1949}%
\bibitem{thibault:1975}%
  \BibitemOpen
  \bibfield{author}{%
  \bibinfo {author} {\bibfnamefont{C.}~\bibnamefont{Thibault}} \emph{et~al.},\
  }%
  \bibfield{journal}{%
  \bibinfo {journal} {Phys. Rev. C}\ }%
  \textbf{\bibinfo {volume} {12}},\ \bibinfo {pages} {644} (\bibinfo {year}
  {1975})%
  \bibAnnoteFile{NoStop}{thibault:1975}%
\bibitem{campi:1975}%
  \BibitemOpen
  \bibfield{author}{%
  \bibinfo {author} {\bibfnamefont{X.}~\bibnamefont{Campi}} \emph{et~al.},\ }%
  \bibfield{journal}{%
  \Doi{http://dx.doi.org/10.1016/0375-9474(75)90065-2}{\bibinfo {journal}
  {Nucl. Phys. A}}\ }%
  \textbf{\bibinfo {volume} {251}},\ \bibinfo {pages} {193} (\bibinfo {year}
  {1975})%
  \bibAnnoteFile{NoStop}{campi:1975}%
\bibitem{warburton:1990}%
  \BibitemOpen
  \bibfield{author}{%
  \bibinfo {author} {\bibfnamefont{E.~K.}\ \bibnamefont{Warburton}}
  \emph{et~al.},\ }%
  \bibfield{journal}{%
  \Doi{http://dx.doi.org/10.1103/PhysRevC.41.1147}{\bibinfo {journal} {Phys.
  Rev. C}}\ }%
  \textbf{\bibinfo {volume} {41}},\ \bibinfo {pages} {1147} (\bibinfo {year}
  {1990})%
  \bibAnnoteFile{NoStop}{warburton:1990}%
\bibitem{otsuka:2001}%
  \BibitemOpen
  \bibfield{author}{%
  \bibinfo {author} {\bibfnamefont{T.}~\bibnamefont{Otsuka}} \emph{et~al.},\ }%
  \bibfield{journal}{%
  \Doi{http://dx.doi.org/10.1103/PhysRevLett.87.082502}{\bibinfo {journal}
  {Phys. Rev. Lett.}}\ }%
  \textbf{\bibinfo {volume} {87}},\ \bibinfo {pages} {082502} (\bibinfo {year}
  {2001})%
  \bibAnnoteFile{NoStop}{otsuka:2001}%
\bibitem{detraz:1979}%
  \BibitemOpen
  \bibfield{author}{%
  \bibinfo {author} {\bibfnamefont{C.}~\bibnamefont{D\'etraz}} \emph{et~al.},\
  }%
  \bibfield{journal}{%
  \Doi{http://dx.doi.org/10.1103/PhysRevC.19.164}{\bibinfo {journal} {Phys.
  Rev. C}}\ }%
  \textbf{\bibinfo {volume} {19}},\ \bibinfo {pages} {164} (\bibinfo {year}
  {1979})%
  \bibAnnoteFile{NoStop}{detraz:1979}%
\bibitem{motobayashi:1995}%
  \BibitemOpen
  \bibfield{author}{%
  \bibinfo {author} {\bibfnamefont{T.}~\bibnamefont{Motobayashi}}
  \emph{et~al.},\ }%
  \bibfield{journal}{%
  \Doi{http://dx.doi.org/10.1016/0370-2693(95)00012-A}{\bibinfo {journal}
  {Phys. Lett. B}}\ }%
  \textbf{\bibinfo {volume} {346}},\ \bibinfo {pages} {9} (\bibinfo {year}
  {1995})%
  \bibAnnoteFile{NoStop}{motobayashi:1995}%
\bibitem{iwasaki:2001}%
  \BibitemOpen
  \bibfield{author}{%
  \bibinfo {author} {\bibfnamefont{H.}~\bibnamefont{Iwasaki}} \emph{et~al.},\
  }%
  \bibfield{journal}{%
  \bibinfo {journal} {Phys. Lett. B}\ }%
  \textbf{\bibinfo {volume} {522}},\ \bibinfo {pages} {227} (\bibinfo {year}
  {2001})%
  \bibAnnoteFile{NoStop}{iwasaki:2001}%
\bibitem{yoneda:2001}%
  \BibitemOpen
  \bibfield{author}{%
  \bibinfo {author} {\bibfnamefont{K.}~\bibnamefont{Yoneda}} \emph{et~al.},\ }%
  \bibfield{journal}{%
  \Doi{http://dx.doi.org/10.1016/S0370-2693(01)00025-9}{\bibinfo {journal}
  {Phys. Lett. B}}\ }%
  \textbf{\bibinfo {volume} {499}},\ \bibinfo {pages} {233} (\bibinfo {year}
  {2001})%
  \bibAnnoteFile{NoStop}{yoneda:2001}%
\bibitem{pritychenko:2002}%
  \BibitemOpen
  \bibfield{author}{%
  \bibinfo {author} {\bibfnamefont{B.~V.}\ \bibnamefont{Pritychenko}}
  \emph{et~al.},\ }%
  \bibfield{journal}{%
  \Doi{http://dx.doi.org/10.1103/PhysRevC.65.061304}{\bibinfo {journal} {Phys.
  Rev. C}}\ }%
  \textbf{\bibinfo {volume} {65}},\ \bibinfo {pages} {061304(R)} (\bibinfo
  {year} {2002})%
  \bibAnnoteFile{NoStop}{pritychenko:2002}%
\bibitem{church:2005}%
  \BibitemOpen
  \bibfield{author}{%
  \bibinfo {author} {\bibfnamefont{J.~A.}\ \bibnamefont{Church}}
  \emph{et~al.},\ }%
  \bibfield{journal}{%
  \Doi{http://dx.doi.org/10.1103/PhysRevC.72.054320}{\bibinfo {journal} {Phys.
  Rev. C}}\ }%
  \textbf{\bibinfo {volume} {72}},\ \bibinfo {pages} {054320} (\bibinfo {year}
  {2005})%
  \bibAnnoteFile{NoStop}{church:2005}%
\bibitem{elekes:2006}%
  \BibitemOpen
  \bibfield{author}{%
  \bibinfo {author} {\bibfnamefont{Z.}~\bibnamefont{Elekes}} \emph{et~al.},\ }%
  \bibfield{journal}{%
  \Doi{http://dx.doi.org/10.1103/PhysRevC.73.044314}{\bibinfo {journal} {Phys.
  Rev. C}}\ }%
  \textbf{\bibinfo {volume} {73}},\ \bibinfo {pages} {044314} (\bibinfo {year}
  {2006})%
  \bibAnnoteFile{NoStop}{elekes:2006}%
\bibitem{gade:2007}%
  \BibitemOpen
  \bibfield{author}{%
  \bibinfo {author} {\bibfnamefont{A.}~\bibnamefont{Gade}} \emph{et~al.},\ }%
  \bibfield{journal}{%
  \Doi{http://dx.doi.org/10.1103/PhysRevLett.99.072502}{\bibinfo {journal}
  {Phys. Rev. Lett.}}\ }%
  \textbf{\bibinfo {volume} {99}},\ \bibinfo {pages} {072502} (\bibinfo {year}
  {2007})%
  \bibAnnoteFile{NoStop}{gade:2007}%
\bibitem{doornenbal:2009a}%
  \BibitemOpen
  \bibfield{author}{%
  \bibinfo {author} {\bibfnamefont{P.}~\bibnamefont{Doornenbal}}
  \emph{et~al.},\ }%
  \bibfield{journal}{%
  \Doi{http://dx.doi.org/10.1103/PhysRevLett.103.032501}{\bibinfo {journal}
  {Phys. Rev. Lett.}}\ }%
  \textbf{\bibinfo {volume} {103}},\ \bibinfo {pages} {032501} (\bibinfo {year}
  {2009})%
  \bibAnnoteFile{NoStop}{doornenbal:2009a}%
\bibitem{huber:1978}%
  \BibitemOpen
  \bibfield{author}{%
  \bibinfo {author} {\bibfnamefont{G.}~\bibnamefont{Huber}} \emph{et~al.},\ }%
  \bibfield{journal}{%
  \Doi{http://dx.doi.org/10.1103/PhysRevC.18.2342}{\bibinfo {journal} {Phys.
  Rev. C}}\ }%
  \textbf{\bibinfo {volume} {18}},\ \bibinfo {pages} {2342} (\bibinfo {year}
  {1978})%
  \bibAnnoteFile{NoStop}{huber:1978}%
\bibitem{keim:2000}%
  \BibitemOpen
  \bibfield{author}{%
  \bibinfo {author} {\bibfnamefont{M.}~\bibnamefont{Keim}} \emph{et~al.},\ }%
  \bibfield{journal}{%
  \bibinfo {journal} {Eur. Phys. J. A}\ }%
  \textbf{\bibinfo {volume} {8}},\ \bibinfo {pages} {31} (\bibinfo {year}
  {2000})%
  \bibAnnoteFile{NoStop}{keim:2000}%
\bibitem{klotz:1993}%
  \BibitemOpen
  \bibfield{author}{%
  \bibinfo {author} {\bibfnamefont{G.}~\bibnamefont{Klotz}} \emph{et~al.},\ }%
  \bibfield{journal}{%
  \Doi{http://dx.doi.org/10.1103/PhysRevC.47.2502}{\bibinfo {journal} {Phys.
  Rev. C}}\ }%
  \textbf{\bibinfo {volume} {47}},\ \bibinfo {pages} {2502} (\bibinfo {year}
  {1993})%
  \bibAnnoteFile{NoStop}{klotz:1993}%
\bibitem{pritychenko:2001}%
  \BibitemOpen
  \bibfield{author}{%
  \bibinfo {author} {\bibfnamefont{B.~V.}\ \bibnamefont{Pritychenko}}
  \emph{et~al.},\ }%
  \bibfield{journal}{%
  \Doi{http://dx.doi.org/10.1103/PhysRevC.63.011305}{\bibinfo {journal} {Phys.
  Rev. C}}\ }%
  \textbf{\bibinfo {volume} {63}},\ \bibinfo {pages} {011305(R)} (\bibinfo
  {year} {2000})%
  \bibAnnoteFile{NoStop}{pritychenko:2001}%
\bibitem{yano:2007}%
  \BibitemOpen
  \bibfield{author}{%
  \bibinfo {author} {\bibfnamefont{Y.}~\bibnamefont{Yano}},\ }%
  \bibfield{journal}{%
  \Doi{http://dx.doi.org/10.1016/j.nimb.2007.04.174}{\bibinfo {journal} {Nucl.
  Instr. Meth. B}}\ }%
  \textbf{\bibinfo {volume} {261}},\ \bibinfo {pages} {1009} (\bibinfo {year}
  {2007})%
  \bibAnnoteFile{NoStop}{yano:2007}%
\bibitem{yoshida:2008}%
  \BibitemOpen
  \bibfield{author}{%
  \bibinfo {author} {\bibfnamefont{A.}~\bibnamefont{Yoshida}} \emph{et~al.},\
  }%
  \bibfield{journal}{%
  \Doi{http://dx.doi.org/10.1016/j.nima.2008.02.05}{\bibinfo {journal} {Nucl.
  Instr. Meth. A}}\ }%
  \textbf{\bibinfo {volume} {590}},\ \bibinfo {pages} {204} (\bibinfo {year}
  {2008})%
  \bibAnnoteFile{NoStop}{yoshida:2008}%
\bibitem{kubo:2003}%
  \BibitemOpen
  \bibfield{author}{%
  \bibinfo {author} {\bibfnamefont{T.}~\bibnamefont{Kubo}},\ }%
  \bibfield{journal}{%
  \Doi{http://dx.doi.org/doi:10.1016/S0168-583X(02)01896-7}{\bibinfo {journal}
  {Nucl. Instr. Meth. B}}\ }%
  \textbf{\bibinfo {volume} {204}},\ \bibinfo {pages} {97} (\bibinfo {year}
  {2003})%
  \bibAnnoteFile{NoStop}{kubo:2003}%
\bibitem{ohnishi:2008}%
  \BibitemOpen
  \bibfield{author}{%
  \bibinfo {author} {\bibfnamefont{T.}~\bibnamefont{Ohnishi}} \emph{et~al.},\
  }%
  \bibfield{journal}{%
  \Doi{http://dx.doi.org/10.1143/JPSJ.77.083201}{\bibinfo {journal} {J. Phys.
  Soc. Jpn.}}\ }%
  \textbf{\bibinfo {volume} {77}},\ \bibinfo {pages} {083201} (\bibinfo {year}
  {2008})%
  \bibAnnoteFile{NoStop}{ohnishi:2008}%
\bibitem{takeuchi:2003}%
  \BibitemOpen
  \bibfield{author}{%
  \bibinfo {author} {\bibfnamefont{S.}~\bibnamefont{Takeuchi}} \emph{et~al.},\
  }%
  \bibfield{journal}{%
  \Doi{http://dx.doi.org/}{\bibinfo {journal} {RIKEN Accel. Progr. Rep.}}\ }%
  \textbf{\bibinfo {volume} {36}},\ \bibinfo {pages} {148} (\bibinfo {year}
  {2003})%
  \bibAnnoteFile{NoStop}{takeuchi:2003}%
\bibitem{agostinelli:2003}%
  \BibitemOpen
  \bibfield{author}{%
  \bibinfo {author} {\bibfnamefont{S.}~\bibnamefont{Agostinelli}}
  \emph{et~al.},\ }%
  \bibfield{journal}{%
  \Doi{http://dx.doi.org/}{\bibinfo {journal} {Nucl. Instr. Meth. A}}\ }%
  \textbf{\bibinfo {volume} {506}},\ \bibinfo {pages} {250} (\bibinfo {year}
  {2003})%
  \bibAnnoteFile{NoStop}{agostinelli:2003}%
\bibitem{mizoi:2005}%
  \BibitemOpen
  \bibfield{author}{%
  \bibinfo {author} {\bibfnamefont{Y.}~\bibnamefont{Mizoi}} \emph{et~al.},\ }%
  \bibfield{journal}{%
  \Doi{http://dx.doi.org/}{\bibinfo {journal} {RIKEN Accel. Progr. Rep.}}\ }%
  \textbf{\bibinfo {volume} {38}},\ \bibinfo {pages} {297} (\bibinfo {year}
  {2005})%
  \bibAnnoteFile{NoStop}{mizoi:2005}%
\bibitem{sakurai:2008}%
  \BibitemOpen
  \bibfield{author}{%
  \bibinfo {author} {\bibfnamefont{H.}~\bibnamefont{Sakurai}},\ }%
  \bibfield{journal}{%
  \Doi{http://dx.doi.org/10.1016/j.nimb.2008.05.090}{\bibinfo {journal} {Nucl.
  Instr. Meth. B}}\ }%
  \textbf{\bibinfo {volume} {266}},\ \bibinfo {pages} {4080 } (\bibinfo {year}
  {2008})%
  \bibAnnoteFile{NoStop}{sakurai:2008}%
\bibitem{nakamura:2009}%
  \BibitemOpen
  \bibfield{author}{%
  \bibinfo {author} {\bibfnamefont{T.}~\bibnamefont{Nakamura}} \emph{et~al.},\
  }%
  \bibfield{journal}{%
  \Doi{http://dx.doi.org/10.1103/PhysRevLett.103.262501}{\bibinfo {journal}
  {Phys. Rev. Lett.}}\ }%
  \textbf{\bibinfo {volume} {103}},\ \bibinfo {pages} {262501} (\bibinfo {year}
  {2009})%
  \bibAnnoteFile{NoStop}{nakamura:2009}%
\bibitem{doornenbal:2010}%
  \BibitemOpen
  \bibfield{author}{%
  \bibinfo {author} {\bibfnamefont{P.}~\bibnamefont{Doornenbal}}
  \emph{et~al.},\ }%
  \bibfield{journal}{%
  \Doi{http://dx.doi.org/10.1016/j.nima.2009.11.017}{\bibinfo {journal} {Nucl.
  Instr. Meth. A}}\ }%
  \textbf{\bibinfo {volume} {613}},\ \bibinfo {pages} {218} (\bibinfo {year}
  {2010})%
  \bibAnnoteFile{NoStop}{doornenbal:2010}%
\bibitem{caurier:2001}%
  \BibitemOpen
  \bibfield{author}{%
  \bibinfo {author} {\bibfnamefont{E.}~\bibnamefont{Caurier}} \emph{et~al.},\
  }%
  \bibfield{journal}{%
  \Doi{http://dx.doi.org/10.1016/S0375-9474(00)00579-0}{\bibinfo {journal}
  {Nucl. Phys. A}}\ }%
  \textbf{\bibinfo {volume} {693}},\ \bibinfo {pages} {374} (\bibinfo {year}
  {2001})%
  \bibAnnoteFile{NoStop}{caurier:2001}%
\bibitem{utsuno:2002}%
  \BibitemOpen
  \bibfield{author}{%
  \bibinfo {author} {\bibfnamefont{Y.}~\bibnamefont{Utsuno}} \emph{et~al.},\ }%
  \bibfield{journal}{%
  \bibinfo {journal} {Nucl. Phys. A}\ }%
  \textbf{\bibinfo {volume} {704}},\ \bibinfo {pages} {50c} (\bibinfo {year}
  {2002})%
  \bibAnnoteFile{NoStop}{utsuno:2002}%
\bibitem{utsuno:1999}%
  \BibitemOpen
  \bibfield{author}{%
  \bibinfo {author} {\bibfnamefont{Y.}~\bibnamefont{Utsuno}} \emph{et~al.},\ }%
  \bibfield{journal}{%
  \Doi{http://dx.doi.org/10.1103/PhysRevC.60.054315}{\bibinfo {journal} {Phys.
  Rev. C}}\ }%
  \textbf{\bibinfo {volume} {60}},\ \bibinfo {pages} {054315} (\bibinfo {year}
  {1999})%
  \bibAnnoteFile{NoStop}{utsuno:1999}%
\bibitem{caurier:1998}%
  \BibitemOpen
  \bibfield{author}{%
  \bibinfo {author} {\bibfnamefont{E.}~\bibnamefont{Caurier}} \emph{et~al.},\
  }%
  \bibfield{journal}{%
  \Doi{10.1103/PhysRevC.58.2033}{\bibinfo {journal} {Phys. Rev. C}}\ }%
  \textbf{\bibinfo {volume} {58}},\ \bibinfo {pages} {2033} (\bibinfo {year}
  {1998})%
  \bibAnnoteFile{NoStop}{caurier:1998}%
\bibitem{poves:1994}%
  \BibitemOpen
  \bibfield{author}{%
  \bibinfo {author} {\bibfnamefont{A.}~\bibnamefont{Poves}} \emph{et~al.},\ }%
  \bibfield{journal}{%
  \Doi{http://dx.doi.org/10.1016/0375-9474(94)90058-2}{\bibinfo {journal}
  {Nucl. Phys. A}}\ }%
  \textbf{\bibinfo {volume} {571}},\ \bibinfo {pages} {221} (\bibinfo {year}
  {1994})%
  \bibAnnoteFile{NoStop}{poves:1994}%
\bibitem{endt:1990}%
  \BibitemOpen
  \bibfield{author}{%
  \bibinfo {author} {\bibfnamefont{P.~M.}\ \bibnamefont{Endt}},\ }%
  \bibfield{journal}{%
  \bibinfo {journal} {Nucl. Phys. A}\ }%
  \textbf{\bibinfo {volume} {521}} (\bibinfo {year} {1990}),\ \doi{\bibinfo
  {doi} {http://dx.doi.org/10.1016/0375-9474(90)90598-G}}%
  \bibAnnoteFile{NoStop}{endt:1990}%
\bibitem{cooper:2002}%
  \BibitemOpen
  \bibfield{author}{%
  \bibinfo {author} {\bibfnamefont{M.}~\bibnamefont{Cooper}} \emph{et~al.},\ }%
  \bibfield{journal}{%
  \Doi{http://dx.doi.org/10.1103/PhysRevC.65.051302}{\bibinfo {journal} {Phys.
  Rev. C}}\ }%
  \textbf{\bibinfo {volume} {65}},\ \bibinfo {pages} {051302(R)} (\bibinfo
  {year} {2002})%
  \bibAnnoteFile{NoStop}{cooper:2002}%
\bibitem{tripathi:2005}%
  \BibitemOpen
  \bibfield{author}{%
  \bibinfo {author} {\bibfnamefont{V.}~\bibnamefont{Tripathi}} \emph{et~al.},\
  }%
  \bibfield{journal}{%
  \Doi{http://dx.doi.org/10.1103/PhysRevLett.94.162501}{\bibinfo {journal}
  {Phys. Rev. Lett.}}\ }%
  \textbf{\bibinfo {volume} {94}},\ \bibinfo {pages} {162501} (\bibinfo {year}
  {2005})%
  \bibAnnoteFile{NoStop}{tripathi:2005}%
\bibitem{tripathi:2006}%
  \BibitemOpen
  \bibfield{author}{%
  \bibinfo {author} {\bibfnamefont{V.}~\bibnamefont{Tripathi}} \emph{et~al.},\
  }%
  \bibfield{journal}{%
  \Doi{http://dx.doi.org/10.1103/PhysRevC.73.054303}{\bibinfo {journal} {Phys.
  Rev. C}}\ }%
  \textbf{\bibinfo {volume} {73}},\ \bibinfo {pages} {054303} (\bibinfo {year}
  {2006})%
  \bibAnnoteFile{NoStop}{tripathi:2006}%
\bibitem{nummela:2001a}%
  \BibitemOpen
  \bibfield{author}{%
  \bibinfo {author} {\bibfnamefont{S.}~\bibnamefont{Nummela}} \emph{et~al.},\
  }%
  \bibfield{journal}{%
  \Doi{http://dx.doi.org/10.1103/PhysRevC.64.054313}{\bibinfo {journal} {Phys.
  Rev. C}}\ }%
  \textbf{\bibinfo {volume} {64}},\ \bibinfo {pages} {054313} (\bibinfo {year}
  {2001})%
  \bibAnnoteFile{NoStop}{nummela:2001a}%
\bibitem{doornenbal:2009}%
  \BibitemOpen
  \bibfield{author}{%
  \bibinfo {author} {\bibfnamefont{P.}~\bibnamefont{Doornenbal}}
  \emph{et~al.},\ }%
  \bibfield{booktitle}{%
  \emph{\bibinfo {booktitle} {Capture Gamma-ray Spectroscopy and Related
  Topics}},\ }%
  \bibfield{journal}{%
  \Doi{http://dx.doi.org/}{\bibinfo {journal} {AIP Conf. Proc.}}\ }%
  \textbf{\bibinfo {volume} {1090}},\ \bibinfo {pages} {639} (\bibinfo {year}
  {2009})%
  \bibAnnoteFile{NoStop}{doornenbal:2009}%
\end{thebibliography}%

\end{document}